\documentclass[12pt]{article}
\usepackage{graphicx}
\begin{document}
\centerline{\bf Computer Simulation of Host and Several Parasite Species with 
Ageing}

\bigskip

Dietrich Stauffer$^{1,3}$ and Karl-Heinz Lampe$^{2,3}$

\bigskip
\noindent

\noindent
$^1$ Institute for Theoretical Physics, Cologne University, 
D-50923 K\"oln, Euroland.

\noindent
$^2$ Zoologisches Forschungsmuseum Alexander Koenig, Adenauerallee 168,
D-53113 Bonn, Germany.

\noindent
$^3$ also IZKS, Bonn University
\bigskip

{\small Abstract: The possible coexistence of one host, one aggressive parasite
and one non-lethal parasite is simulated using the Penna model of biological 
ageing. If the aggressive parasites survive the difficult initial times 
where they have to adjust genetically to the proper host age, all three 
species may survive, though the host number may be diminished by 
increasing parasite aggressivity. Simulations of more
diversified conditions considering more than two parasite species underline
the importance of synchronization for long term survival.}

\bigskip

The extinction of biological population is an extreme event (Albeverio et al
2006) and in an ageing simulation like the Penna model can be due to random 
fluctuations for small populations (Pal 1996) or 
due to unfavourable conditions for large populations (Malarz 2007). We had 
earlier applied these computer simulation methods to one host species and one 
parasite species, where the parasite needs to attack the host at one specific 
age of the host (Stauffer et al. 2007). Now we generalise this simulation 
to two parasite species, and assume that one of them also kills the host with 
some probability and the other (non-lethal) parasite with some other probability
(Lampe 1984).

All these simulations were based on the at present most commonly used computer 
model of biological ageing, the Penna model (Penna 1995, Stauffer et al. 2006) 
of Medawar's mutation accumulation; see e.g. Mueller and Rose (1996) and 
Charlesworth (2001) for alternative implementations of mutation accumulation. 
The genome is represented by 32 bits, where a zero bit means healthy and 1
means sick. The position of a 1-bit in this bit-string gives the age interval
from which on a life-threatening disease affects the health. If three such 
diseases are active (3 bits set to 1 up to that age), the individual dies at
the age corresponding to the position of the third 1-bit. In addition, all 
individuals die independently of their age with the Verhulst probability $N(t)/K$
at time step $t$, where $N$ is the current population size and $K$ is often 
called the carrying capacity. At each iteration, each living mature individual 
of age 8 and above gives birth to $B$ offspring, with $B=3$ for the host and 
$B=14$ for the parasite. At each
birth, one random mutation is made for all offspring by flipping a randomly
selected bit from 0 (healthy) to 1 (sick). (If it is 1 already it stays at 1:
no new mutation.) Typically, $10^4$ iterations were 
averaged over after the populations had reached roughly a dynamic equilibrium 
of births and deaths. More details on the Penna model of 1995 are given in many 
articles and books (Stauffer et al. 2006, Stauffer 2007) and in our appendix. 
[For example, this Penna model predicted (Altevolmer 1999) some 
counterintuitive (Mitteldorf and Pepper 2007) effects of predators on 
ageing (Reznick et al 2004).]

\begin{figure}[hbt]
\begin{center}
\includegraphics[angle=-90,scale=0.5]{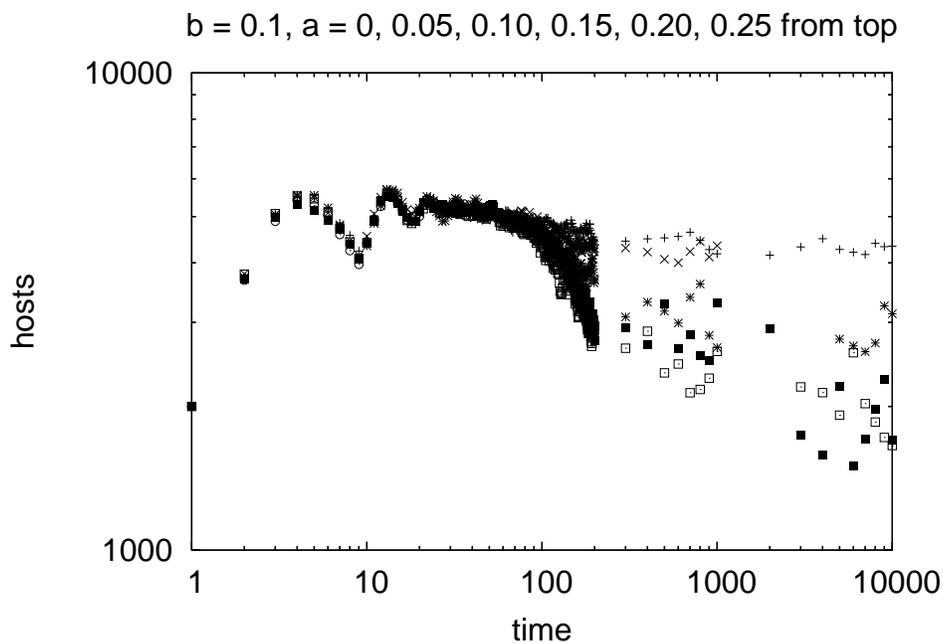}
\end{center}
\caption{Time dependence of host number, for fixed aggressiveness $b$ of 
parasite P1 against the other parasite P2, and for varying aggressiveness
$a$ of P1 against the host.
}
\end{figure}

\begin{figure}[hbt]
\begin{center}
\includegraphics[angle=-90,scale=0.5]{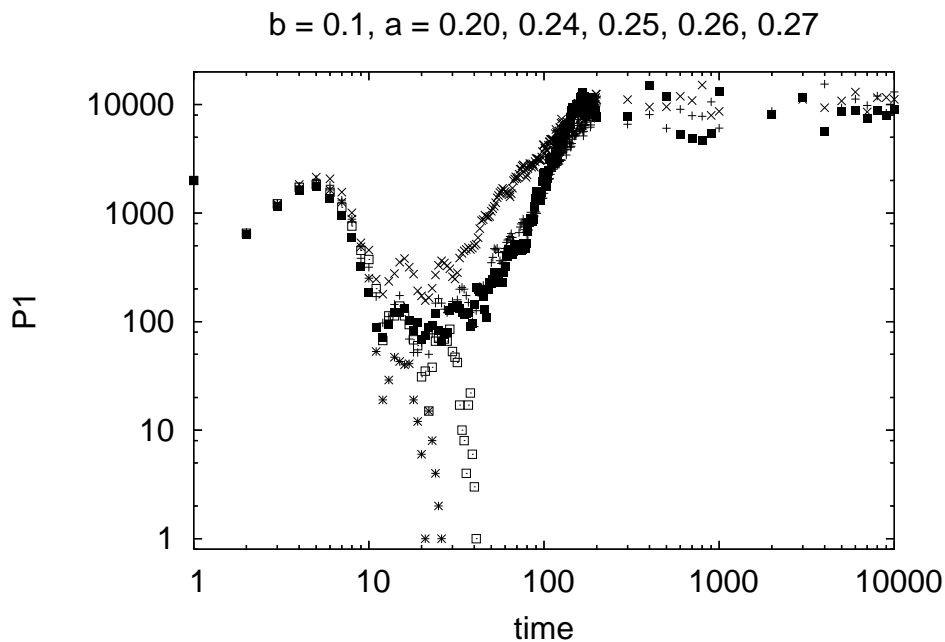}
\end{center}
\caption{As previous figure but for the number of aggressive parasites P1.
}
\end{figure}

\begin{figure}[hbt]
\begin{center}
\includegraphics[angle=-90,scale=0.5]{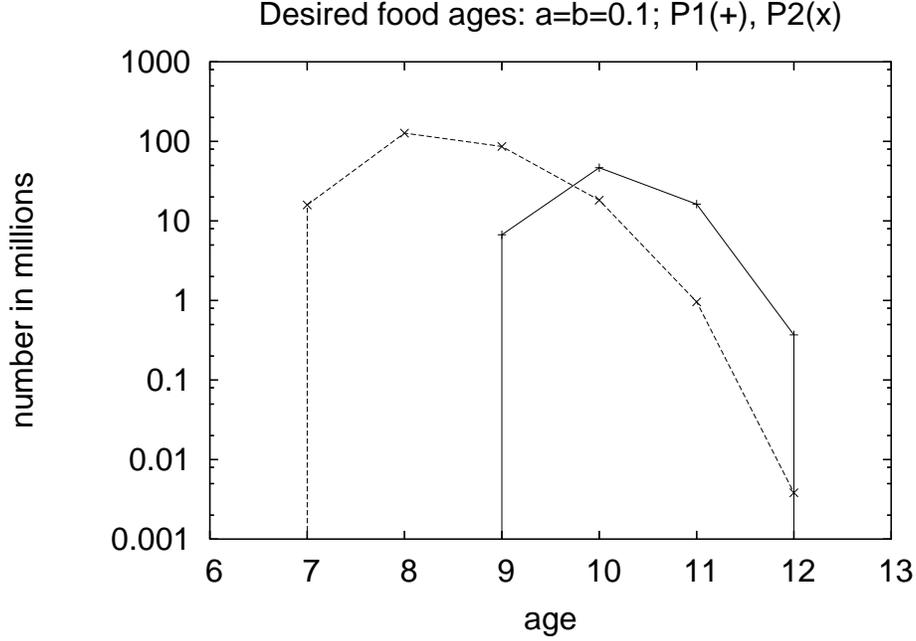}
\end{center}
\caption{Final distribution of genetically programmed desired host ages
for the two parasite species P1 (right) and P2 (left). The initial distribution 
is flat, between ages 1 and 31.
}
\end{figure}

\begin{figure}[hbt]
\begin{center}
\includegraphics[angle=-90,scale=0.5]{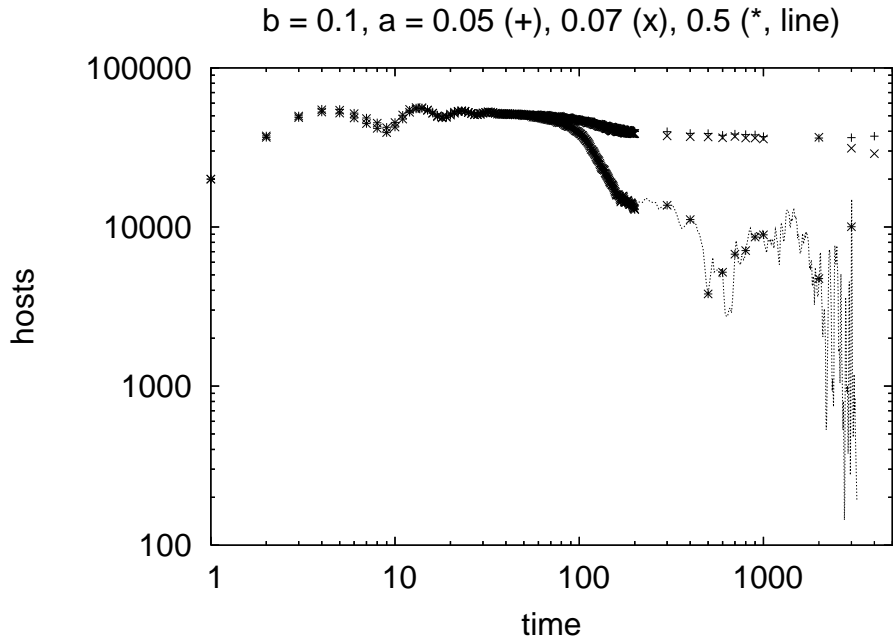}
\end{center}
\caption{Ten times bigger initial populations (of all three species) than in 
Figs.1,2 avoids the parasite P1
extinction seen in some cases of Fig.2 at short times. The 
line for $a = 0.5$ shows the enormous later fluctuations in the host numbers.
}
\end{figure}

\begin{figure}[hbt]
\begin{center}
\includegraphics[angle=-90,scale=0.45]{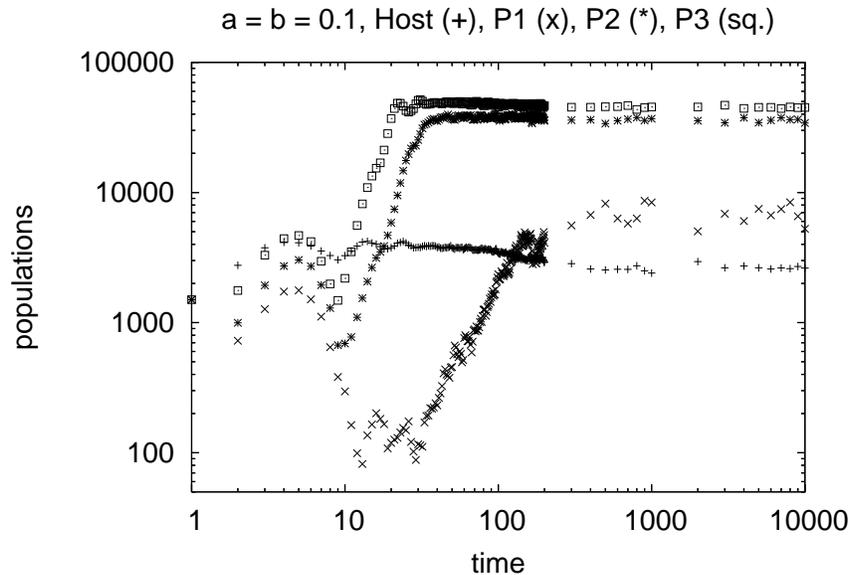}
\end{center}
\caption{Example with one lethal and two nonlethal parasite species; P1 
can kill both non-lethal species with the same probability $b$. For $a=0$ the 
results look similar.
}
\end{figure}

\begin{figure}[hbt]
\begin{center}
\includegraphics[angle=-90,scale=0.3]{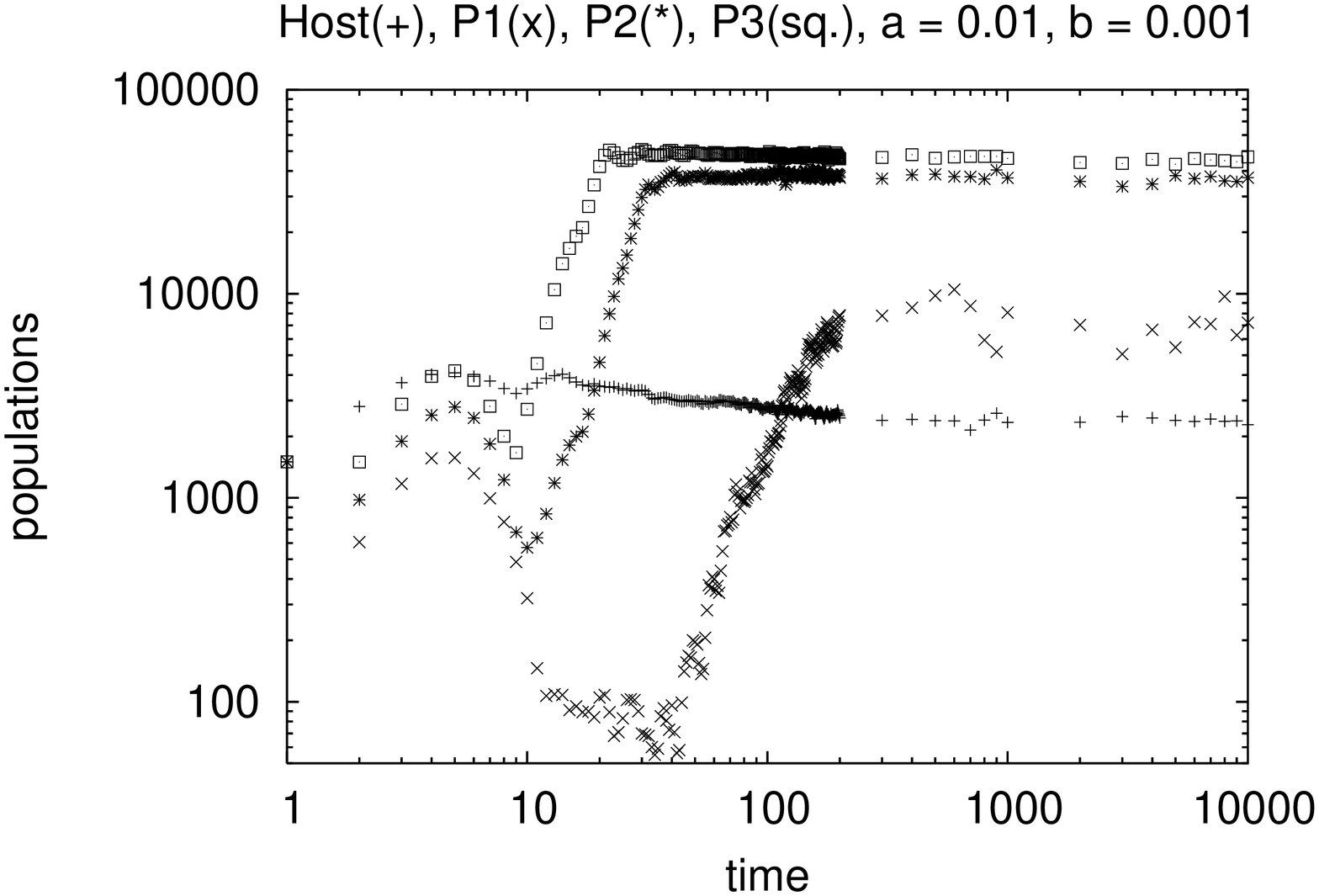}
\includegraphics[angle=-90,scale=0.3]{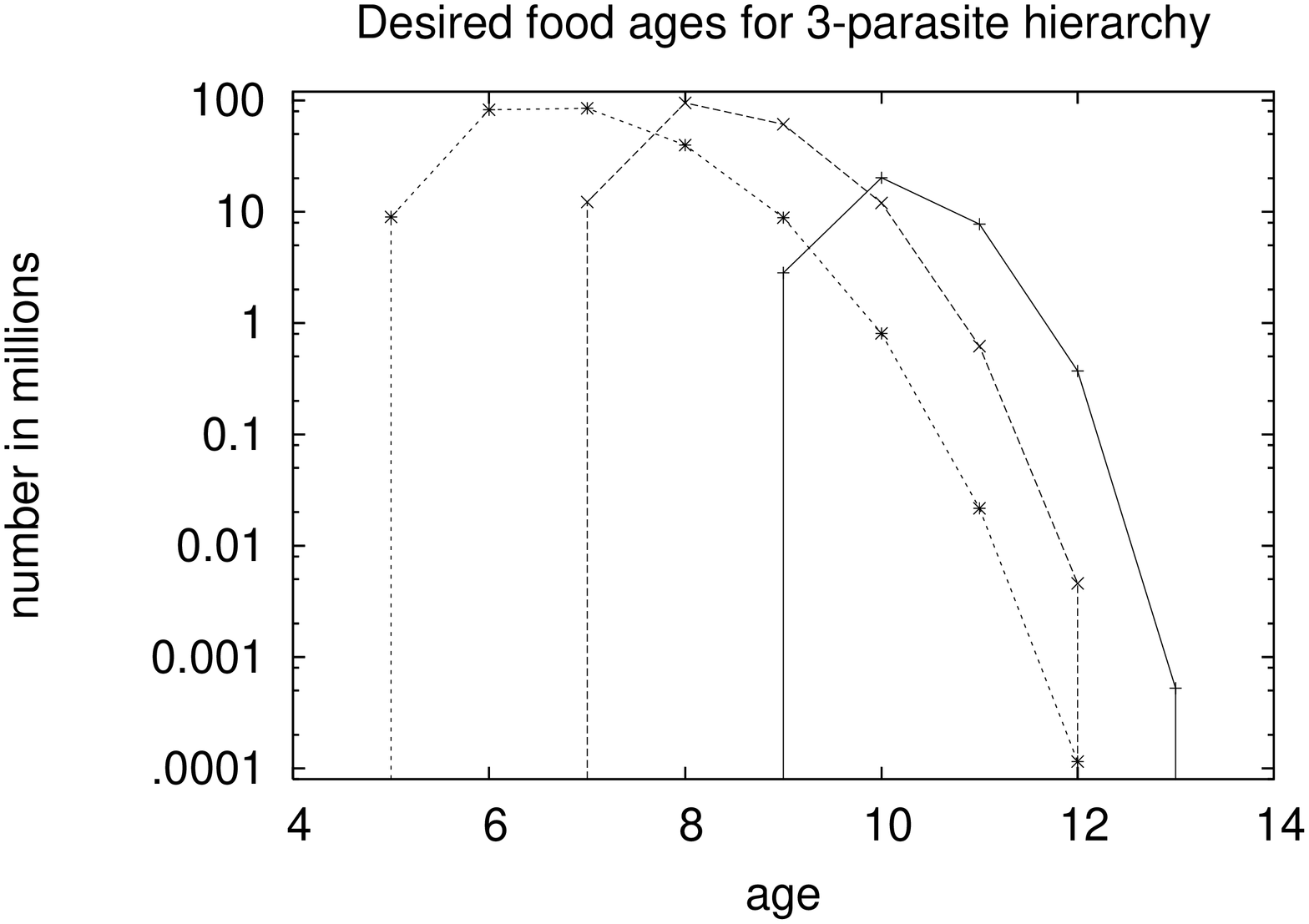}
\end{center}
\caption{Hierarchy of three parasites. Part a shows an example of coexistence
and part b for this example the distribution of the desired host ages for the
three parasite species. 
}
\end{figure}
\begin{figure}[hbt]
\begin{center}
\includegraphics[angle=-90,scale=0.28]{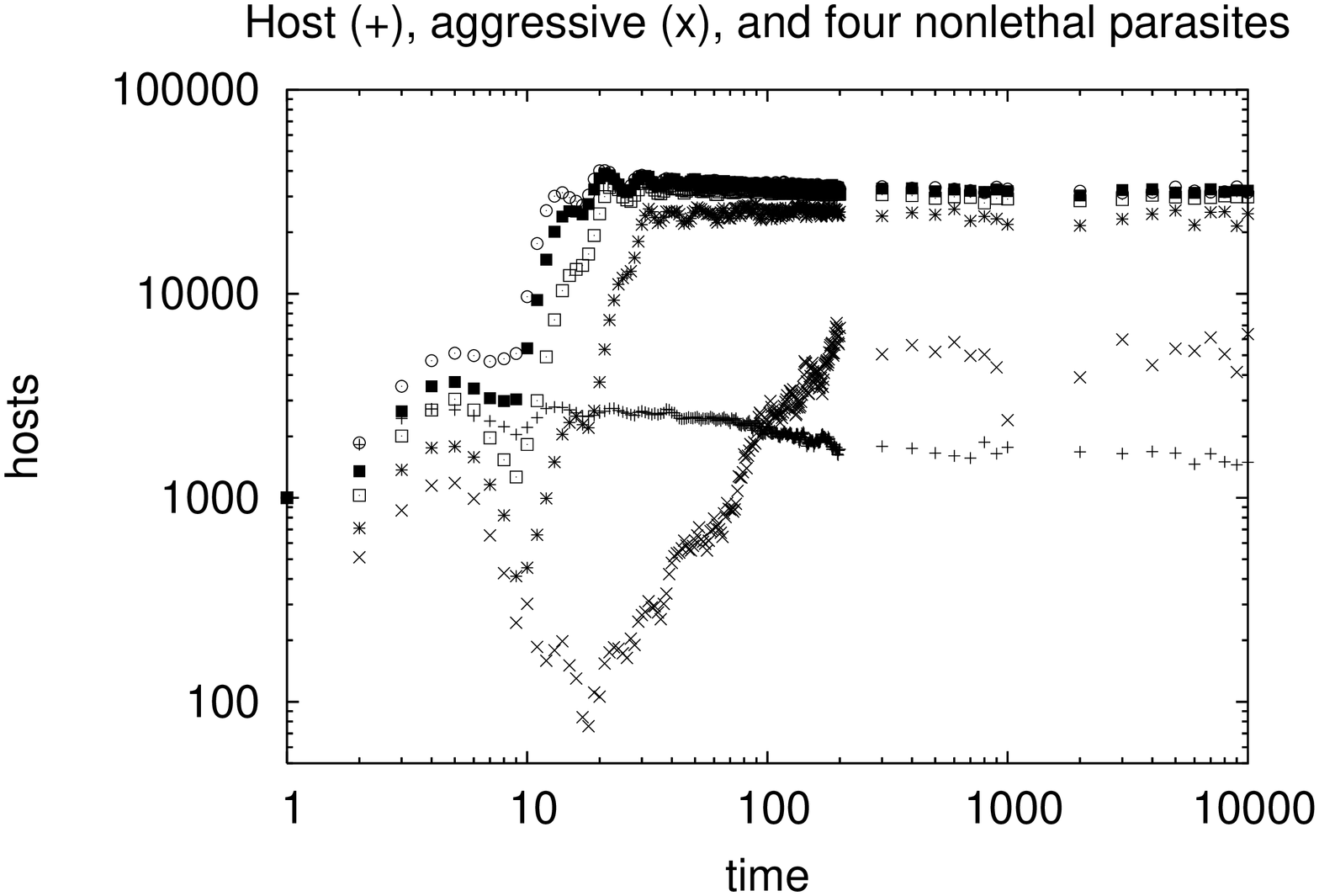}
\includegraphics[angle=-90,scale=0.28]{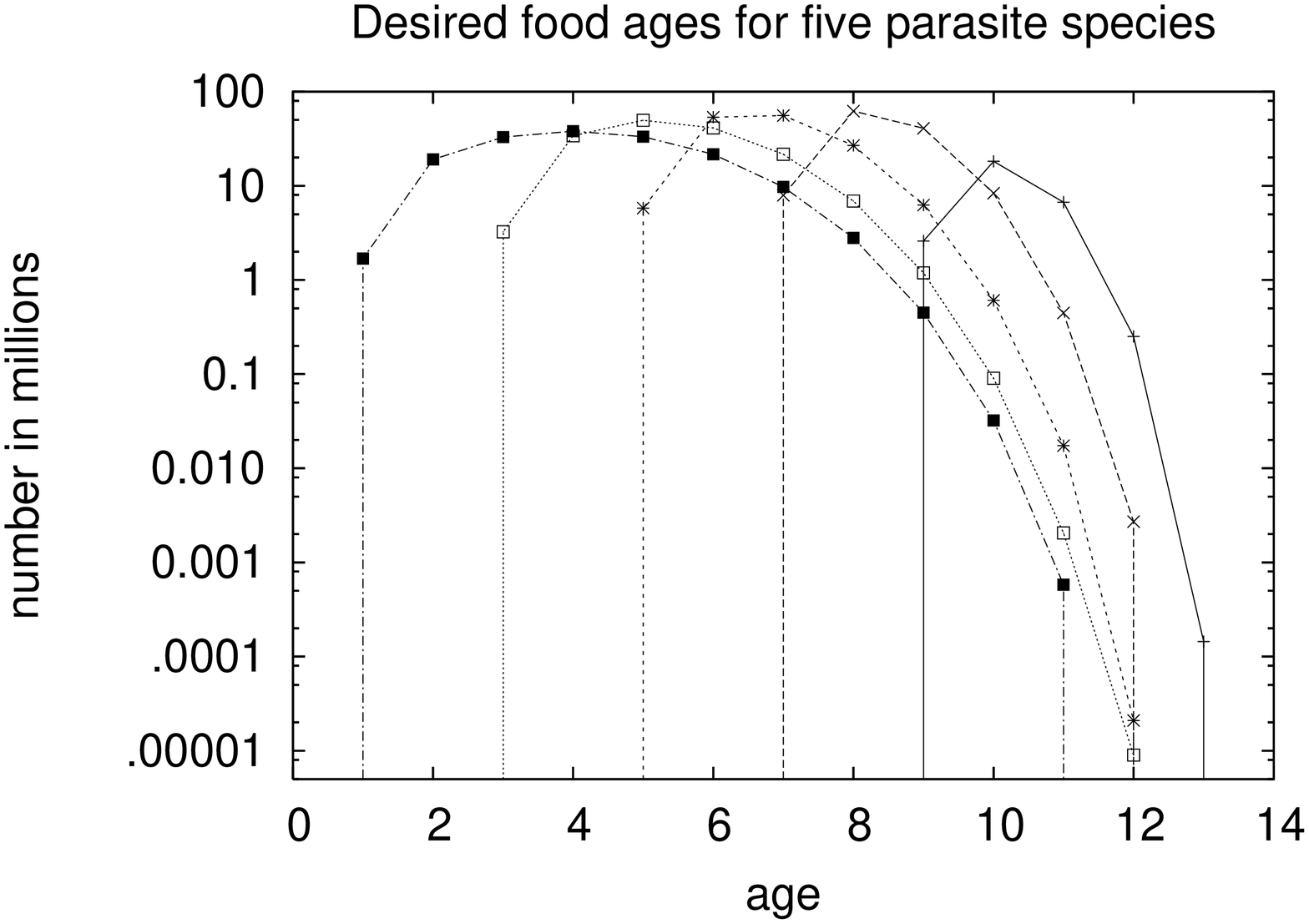}
\end{center}
\caption{Part a: As Fig.6 but for one aggressive (x) and four non-lethal 
parasites attacking the host(+). Part b shows the synchronization of the five
parasite species, with minimum food ages 2,4,6,8,10 from left to right.
}
\end{figure}

\begin{figure}[hbt]
\begin{center}
\includegraphics[angle=-90,scale=0.34]{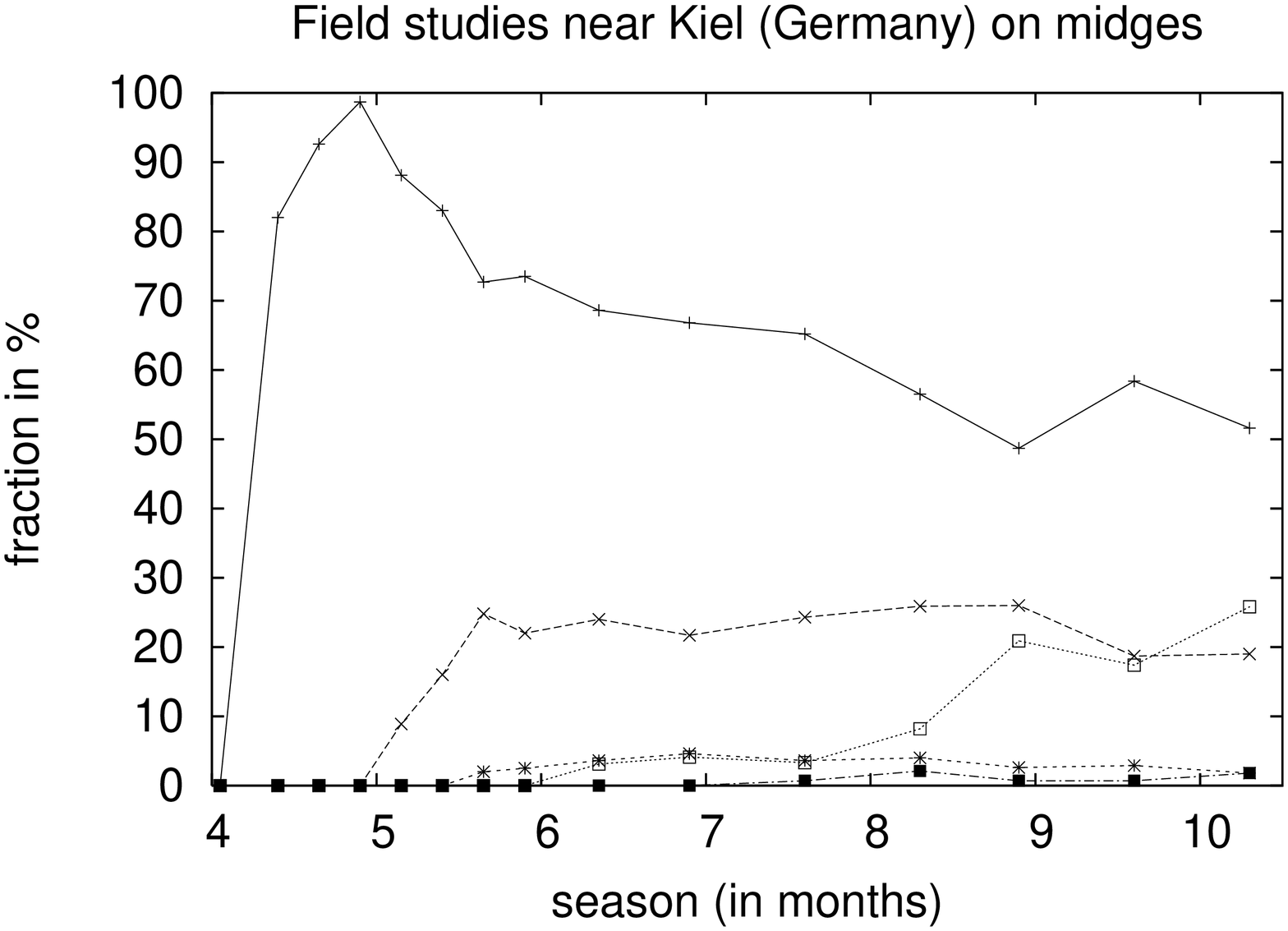}
\includegraphics[angle=-90,scale=0.34]{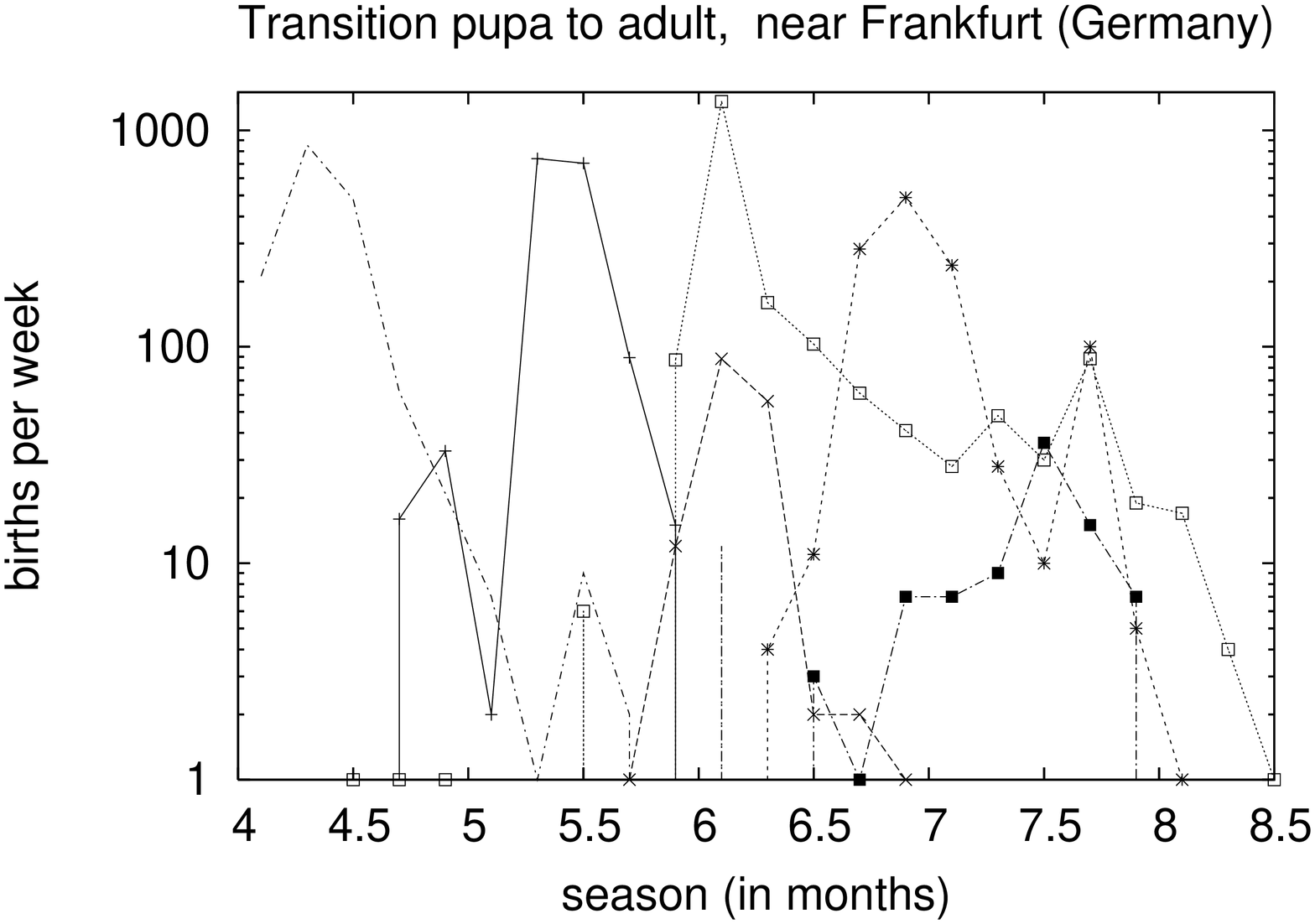}
\end{center}
\caption{Top: Fraction of {\it Mikiola fagi} Hartig, 1839 hosts infected by 
five different insect species versus months (May to November 1984)
of the year: {\it Omphale lugens} (Nees, 1853) (+), {\it Aprostecetus
elongatus} (F\"orster, 1841) (x), {\it Torymus cultiventris} Ratzeburg 1844
(*), {\it Mesopolobus fagi} Askew \& Lampe, 1998
(full squares), {\it Aprostecetus lycidas} (Walker, 1839) (empty squares).
Bottom: Number of emerging adults for {\it O. 
lugens} (+),  {\it A. elongatus} (x), {\it Eupelmus urozonus} Dalman, 1820
(*), {\it Eumacepulus grahami} von Rosen, 1960 (full squares), 
{\it A. lycidas} (empty squares) and the host {\it M. fagi} (leftmost line). 
For {\it Aprostecetus luteus} (Ratzeburg, 1852) the maximum was at 
8.7 months, for {\it Torymus fagi} (Hoffmeyer, 1930) at 9.6 months (not shown).
}
\end{figure}

Many other mathematical or computational studies of host-parasite systems and 
their (co-)evolution were published (see e.g. van Baalen and Sabelis 1995,
Martins 2000, Restif and Koella 2003, Tseng 2006) but most do not include the
individual ages, while our aim is the emergence of age synchronisations. Only 
Martins 2000 included ageing, using the same Penna model as we do but applying it
to a comparison of sexual versus asexual host reproduction in the presence of
parasites. 

With this standard model, hosts and parasites are simulated together, using 
separate Verhulst factors with the carrying capacity of the parasites ten 
times bigger than that of the host. In contrast to the hosts, the parasites at 
each iteration make 100 attempts to invade a living host of their desired age. 
(These desired host ages are at the beginning distributed randomly
between zero and 31). If all 100 attempts were unsuccessful, they die. Otherwise
they have enough food provided the host age is at least 10 for the first 
(aggressive) species of parasite called P1, and is at least 8 for the second 
non-lethal species of parasite called P2. If the host is too young, then again 
the parasite dies; if not it survives and ages. 

We also mutated, randomly and continuously, the desired host age stored in the
genome of each parasite. Then at birth of a parasite, not only the usual 
bit-string genome is mutated but also its desired host age. With 25 percent 
probability it increases by one unit, with 25 percent probability it decreases
by one unit, and in the remaining 50 percent of the cases it stays constant.
Initially we have equally strong populations of host and each parasite species.

Let $i=1, 2, 3$ denote the aggressive parasites, the non-lethal parasites, 
and the hosts, $K_i$ with $K_1 = K_2 = 10K_3$ the carrying capacities, and
$K_3$ is ten times the initial population $N_i(t=0)$ (same for each species).
Then the three Verhulst death probabilities are assumed as
$$ p_1 =   N_1 /K_1 $$
$$ p_2 = (N_2+bN_1)/ K_2 $$
$$ p_3 = (N_3+aN_1)/ K_3 $$
with $b = 0.1$ and various aggressivities $a$ in our simulations. Thus only
the aggressive parasite has the usual Verhulst factor for $N_1$, while both the
host number $N_3$ and the number $N_2$ of non-lethal parasites are 
negatively influenced by the $N_1$ aggressive parasites.

Figure 1 shows the time dependence of the host population and
Fig.2 that of P1. The one for P2 is not shown since it barely changes when we 
change parameters and always survived. Fig.3 shows that the distributions
of the desired host ages have self-organized to rather sharp peaks centered 
about the given minimum host ages of 10 for P1 and 8 for P2.

Small populations subject to random fluctuations always die out if observed long
enough (Pal 1996). ``Small'' means here that the fluctuations in the population
size are much smaller than the average population size. This effect is 
responsible for the extinction of P1 seen in some of the curves of Fig.2. With 
a ten times bigger population, the time consuming simulations of Fig.4 show
survival of P1 were before we had extinction: Instead of extinction we now 
see a minimum. These surviving parasites P1 later reduce the host population. 

We also simulated three instead of two parasite species, adding a non-lethal
P3 feeding on host of age 6 and above, compared with 10 and 8 for P1 and P2. 
Fig.5 shows an example of coexistence; in another sample differing only in the 
random numbers the parasite P1 became extinct. The analog of Fig.3 now gives
three peaks (not shown.) 

More complex is a system of three aggressive parasites P1, P2, P3 plus one host: 
each parasite species $i$ kills the host with probability $\propto aN_i$, while
parasites kill each other with probability $\propto bN_i$ in a hierarchical way:
P3 kills P1 and P2, P2 kills P1. Usually one of the species dies out but Fig.6
shows an example of coexistence for small $b = 0.001, \; a = 0.01$ and again minimum
food ages 10, 8 and 6 for P1, P2 and P3, respectively. The three self-organised
distributions of desired food ages show again nicely the importance of 
synchronisation.

Finally we simulate one aggressive and four non-lethal parasite species, with
needed host ages of 10, 8, 6, 4 and 2, respectively, in Fig.7. Fig.8 corresponds
to reality, as observed near Kiel in northern Germany among 200 to 450
hosts of \textit{Mikiola fagi} (Hartig 1839) (European beech tree gall midge), 
and five parasite species attacking at different times of the year. This 
synchronisation of lives is what we simulated here.

In summary, for three to six species as well as in the previous study 
(Stauffer et al. 2007)
of only one type of parasites, the proper distribution of desired host ages 
emerges in the parasite genomes, provided the aggressive parasites do not go 
extinct during this process of self-organisation.

\bigskip
We thank K. Rohde for drawing our attention to this journal and its symbiosis
issue, and for a critical reading of the manuscript.

\bigskip
{\bf References}
\parindent 0pt

\medskip
Albeverio S, Jentsch V, Kantz H  eds (2006). 
Extreme Events in Nature and Society,  Springer, Berlin-Heidelberg.

\medskip
Altevolmer AK (1999). Virginia opossums, minimum reproduction age and predators
in the Penna aging model. Int. J. Modern Physics. C 10: 717-722.

\medskip
Charlesworth B (2001).  Patterns of age-specific means and genetic variances of 
mortality rates predicted by the mutation-accumulation theory of ageing. J.
Theor. Biol. 210: 47-65.

\medskip
Lampe K-H (1984). Struktur und Dynamik des Parasitenkomplexes der
Binsensacktr\"agermotte Coleophora alticolella Zeller (Lep.:
Coleophoridae) in Mitteleuropa. - Zool. Jb. Syst. 111: 449-492.

\medskip
Malarz K (2007). Risk of extinction - mutational meltdown or the 
overpopulation, Theory in Bioscience 125: 247-156.

\medskip
Martins JS S\'a (2000). Simulated coevolution in a mutating ecology. Physical
Review 61: 2212-2215.

\medskip
Mitteldorf J, Pepper JW (2007). How can evolutionary theory accommodate 
recent empirical results on organismal senescence? Theory in Biosciences
126: 3-8.

\medskip
Mueller, LD, Rose, MR (1996). Evolutionary theory predicts late-life mortality 
plateaus. Proc. Natl. Acad. Sci. USA 93: 15249-14253.

\medskip
Pal KF (1996). Extinction of small populations in the bit-string model
of biological ageing. Int. J. Modern Physics C 7: 899-908.

\medskip
Penna TJP (1995). A bit-string model for biological aging. J. 
Statistical Physics 78: 1629-1633.

\medskip
Restif, O; Koella, JC (2003). Shared control of epidemiological traits in a 
coevolutionary model of host-parasite interactions. Am. Nat. 161: 827-836.

\medskip
Reznick DN, Bryant MJ, Roff D, Ghalambor CK, Ghalambor DE (2004). Effect of 
extrinsic mortality on the evolution of senescence in guppies. Nature 431:
1095-1099.

\medskip
Stauffer D, Moss de Oliveira S, de Oliveira PM, S\'a Martins JS (2006).
{\it Biology, Sociology, Geology by Computational Physicists}.  Elsevier,
Amsterdam.

\medskip
Stauffer D, Proykova A, Lampe K-H (2007). Monte Carlo 
Simulation of age-dependent host-parasite relations, Physica A, in press.
doi:10.1016/j.physa. 2007.05.028

\medskip
Stauffer D (2007). The Penna model of biological aging. Bioinformatics and 
Biology Insights. 1: 91 (2007) (electronic only: http://la-press.com )

\medskip
Tseng M (2006). Interactions between the parasite's previous and current environment mediate the outcome of parasite infection. Am. Nat. 168: 565-571.

\medskip
van Baalen M, Sabelis MW (1995). The dynamics of multiple infection and the
evolution of virulence. Am. Nat. 146: 881-910.

\end{document}